# Effect of Graded Bias Voltage on the Microstructure of arc-PVD CrN Films and its Response in Electrochemical & Mechanical Behavior

M.L. Cedeño-Venté[a], D.G. Espinosa-Arbeláez[a], J. Manríquez-Rocha[b], G.C. Mondragón Rodríguez[a], A.E. Gómez-Ovalle[a], J. González-Hernández[a], J.M. Alvarado-Orozco[a]

[a] Centro de Ingeniería y Desarrollo Industrial, Surface Engineering Dept., Av. Pie de la Cuesta 702, 76125 Santiago de Querétaro México.
[b] Centro de Investigación y Desarrollo Tecnológico en Electroquímica, Pedro Escobedo, 76730 Santiago de Querétaro México.

**Abstract**

The effect of graded or constant bias voltages (-40 V, -80 V and -40/60/80 V) on size grain and surface defects of arc PVD deposited CrN films was investigated. Corrosion resistance evaluated using electrochemical impedance spectroscopy (EIS) and potentiodynamic curves (Tafel) and the mechanical behavior evaluated by means of instrumented nanoindentation and scratch testing was correlated with the microstructural changes. It was found that the bias voltage variation affects corrosion behavior due to the presence of defects (i.e. open voids, droplets) which also affects the failure mechanisms and increasing spallation. High bias voltage (-80 V) increases nano-hardness and the elastic modulus due to the dense microstructure of the CrN coating.

1. **Introduction**

Chromium nitride coatings deposited by physical vapor deposition technologies (PVD) have been wide interest owing to the ability to improve corrosion, oxidation and wear resistance; as well as the low friction coefficient [1]. For those reasons, it is a suitable candidate for uses in mechanical devices, cutting and forming tools, die casting molds and plastic extrusion [2]. Some PVD technology variants such as cathodic arc (arc-PVD) and magnetron sputtering (MS) have been widely used to deposit CrN. Particularly, the cathodic arc has gained importance for the high deposit speeds and excellent substrate adhesion [3]. However, during the deposition process, unavoidable defects are generated which reduce the mechanical properties and the corrosion resistance of the coatings. These defect such as open voids, pinholes, cone-like defects and droplets act as ways for the development and propagation of corrosion phenomenon. For example, the pinholes derived from the shadowing effect and the voids caused by the removal of particles weakly bound to the coating can act as routes of diffusion of corrosive media towards the substrate, allowing the degradation and subsequent delamination [4], also it reduces the densification of the film, which leads to a decrease in the mechanical performance of the piece. On the other hand, droplets generated by molten cathode material can decrease coating adhesion, increasing the roughness and porosity or acting as abrasive particles in the coating [5]. Additionally, these defects can accelerate the substrate corrosion by the generation of localized galvanic corrosion [6,7]; however, their effects are sometimes controversial [8][9].

To avoid the droplets inconvenience, various plasma filtering techniques have been developed, but the up-scalability of many filters is not easy to get due to their geometry. Additionally, the effect of process parameters has been considered as a good way for getting this aim. An appropriate control of parameters as cathode voltage, substrate temperature, gases ratio, bias voltage improve coatings performance and they can reduce defects density. In CrN films an increase in $N_2$ pressure produce phase's transformation, changes in preferential orientation and it can increase macroparticles size and numbers [10]. The bias voltage, which is applied to the substrate affects surface morphology, the mechanical and electrochemical properties of films. Constant voltages higher reduce size and number of droplets, this phenomenon can be originated for two reasons: 1) a higher bias voltage produces the accelerated particles have higher momentum and more macroparticles are removed of the coating. 2) The electrical repulsion force due to change the charging state of the macroparticles the higher the bias voltage [11].

Recently, graded bias voltage has been implemented to find deposition parameters that assure improved coating properties, however, there are few studies about graded bias in nitride films and its effect on microstructure and corrosion resistance of coatings. Shuyong Tan [12] found the hardness of bias-graded films is between that of films with the low and high constant voltage. Also, the graded bias voltage leads to the modification of the morphology and microstructure of the CrN film, moving from a pyramidal or cellular conical structure to a mixture of both.

In this work, we study the influence of constant and graded bias voltage on the microstructure and formation of surface defects and their relationship with electrochemical behavior and mechanical in CrN coating systems.

## 2. Experimental Methodology

### 2.1. Coatings deposition

CrN thin films were deposited by arc sputtering using Oerlikon Balzers PVD equipment, model DOMINO MINI MZR 303/4/46. Chromium targets with 99.95% of purity were used. Substrates M2 steel were polished to a roughness of Ra 0.025 micrometer and they were chemically cleaned with acetone. Finally, the cleaning was finished with an ions cleaning process into the chamber using a working pressure of $1 \times 10^{-2}$ mbar and bias voltage -250 V for 30 min. A reactive nitrogen discharge using a flow of 500 sccm was used to deposit CrN coatings. The current and temperature of the substrate were kept at 150 A and 400 °C, respectively. To study the effect of the bias voltage, the substrates were polarized with constant voltages of -40V and -80V. A graded bias voltage of -40/-60/-80V was also applied. The remaining experimental conditions were maintained constant for all samples.

### 2.2. Characterization of films

*Microstructure, morphology, and chemistry*

The crystallite size and crystal phases of the samples were analyzed by X-ray diffraction (XRD) using a SmartLab diffractometer (Rigaku), with a Cu Kα radiation (λ = 0.15405 nm). Grazing incidence scans in the range 30°– 90° at 1° incidence, with a step size of 0.02° were used.

The coating thickness and composition as a function of the deep were measured with glow discharge optical emission spectrometry (GD-OES, Horiba) equipment which was operated at 650 Pa and 35W of pressure and power, respectively. The GD-OES data was complemented by a scanning electron microscope (SEM) and energy dispersive spectroscopy (EDS).

The samples microstructure was studied in a Helios Nanolab 600 SEM using an electron secondary detector at 5 kV of accelerating voltage. The surface roughness of coated and uncoated specimens was studied using a Dektak Xt-A surface profilometer (Bruker, Billerica, MA, USA) with a conical tip of 2 µm radius and a scan rate of 66 µm/s.

Optical microscopy was carried out at the surface of the specimen using a VHX-5000 digital (Keyence). Images at 500X magnifications were used to determine the number and average size of defects. Optical images with the same properties were analyzed by ImageJ software program and a complementary analysis with 3D profilometry images was made on 500µm x 500µm.

*Mechanical properties*

A scratch tester (Anton Paar) with a 200 µm diamond tip was used to study the film adhesion. The tests were carried out with progressive loads from 1 N to 35 N at a loading speed of 68 N/min. The cohesive and adhesive critical loads were identified through the correlation between the signal of an acoustic sensor and the images of the scratch tracks. Failure mechanisms were identified and studied by scanning electron microscopy.

Oliver and Pharr method was used to evaluate the hardness and reduced elastic modulus using a Nanoindentation Hysitron TI-700 UBI 1 with a Berkovich (BK2) indenter at a normal load of 6 mN for 10 s. 60 nanoindentations were performed per sample and Weibull statistics was used to analyze the experimental data.

*Corrosion testing*

Corrosion tests were carried out using a BioLogic potentiostat-galvanostat SP300. A typical three-electrode cell was used with a platinum mesh counter electrode and an Ag/AgCl reference electrode. The coated and uncoated M2 steels were employed as a working electrode with a 0.95 cm² working area. The electrolyte consisted of a 3.5% NaCl solution in deionized water at room temperature (25 °C).

Potentiodynamic polarization tests were obtained with 2 of mV/s scan rate and potentials between -1 V to +1 V with respect to the open circuit potential. The Ecorr parameter, anodic (βa) and cathodic (βc) slopes, and the corrosion current density (icorr) were extracted from Tafel plots. The protective efficiency (Pe) was determined from the corrosion current densities of the coatings ($I_{corr}$) and the substrate ($I_{corr}^0$).

$$P_e(\%) = \left(1 - \frac{I_{corr}}{I_{corr}^0}\right) \times 100 \qquad (1)$$

## 3. Results and discussion

### *Microstructure, morphology and chemistry*

Figure 1 shows XRD patterns of CrN coatings which exhibit diffraction peaks corresponding to the B1 type structure of CrN phase (ICSD card # 01-083-5612, in figure 1a) with strong reflection in (200), as can be seen in figure 1b. The samples were compared with previously reported of unstrained CrN and they are presented in table I. A compressive stress state are present in the samples under study which are according to the shifting (200) peak to higher angles observed for $40V_b$, $40/60/80V_b$, and $80V_b$ samples. This behavior is related to the residual stresses due to atomic peening effect (APE) during the film growth [13].

A higher lattice strain was found from the 80Vb specimen caused by APE that promotes changes in nucleation and growth kinetics. Table I shows that crystallite size increases with the corresponding increment of the bias voltage. It is due to the enhanced mobility of adatoms on the surface and hence the growth and coalescence of nuclei were accelerated [14]. Likewise, it was observed a correlation between Bias Voltage and coating thickness finding a less thickness and major density of the coating for $80V_b$ [15].

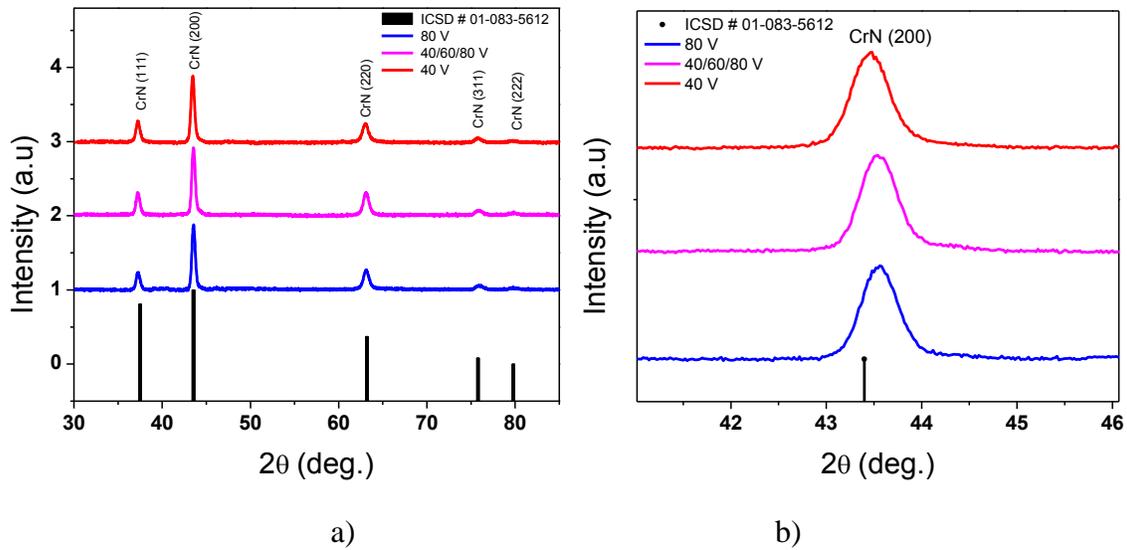

*Figure 1. XRD patterns of CrN films: a) at 40 $V_b$, 40/60/80 $V_b$ and 80 $V_b$, and b) zoom in of (200) peak.*

The GDOES measurements show the N/Cr ratio in the CrN films was reduced with the increase in negative bias voltage [15]. The reduction of N/Cr ratio was related to the re-sputtering phenomenon of the N atoms which is originated when the ions impinging increase their energy. The high-energy ions selectively sputter the N atoms because they are lighter than Cr atoms.

*Table I. Crystallographic parameters and structure properties*

| Specimen | Lattice strain (%) | Lattice constant (Å) | Crystallite size (Å) | Thickness (µm) | N/Cr (%wt) |
|---|---|---|---|---|---|
| **40V$_b$** | 35.000 | 4.1660 | 179 | 5.060 | 0.166 ± 0,004 |
| **40/60/80V$_b$** | 39.000 | 4.1637 | 220 | 5.010 | 0.219 ± 0,004 |
| **80V$_b$** | 60.000 | 4.1656 | 232 | 4.020 | 0.102 ± 0,001 |
| **ICSD card # 01-083-5612** | Unstrained | 4.1668 | - | - | - |

Figure 2 shows the roughness values obtained by profilometry. The films roughness and defects percent have a clear tendency to be reduced for coatings obtained with a higher bias voltage; such as is previously reported by Warcholinski *et al.* who found a decrease of the microparticles density when the bias voltage increase [17]. However, we found an increase in the particle average size that could be related to higher arcs formation during the process.

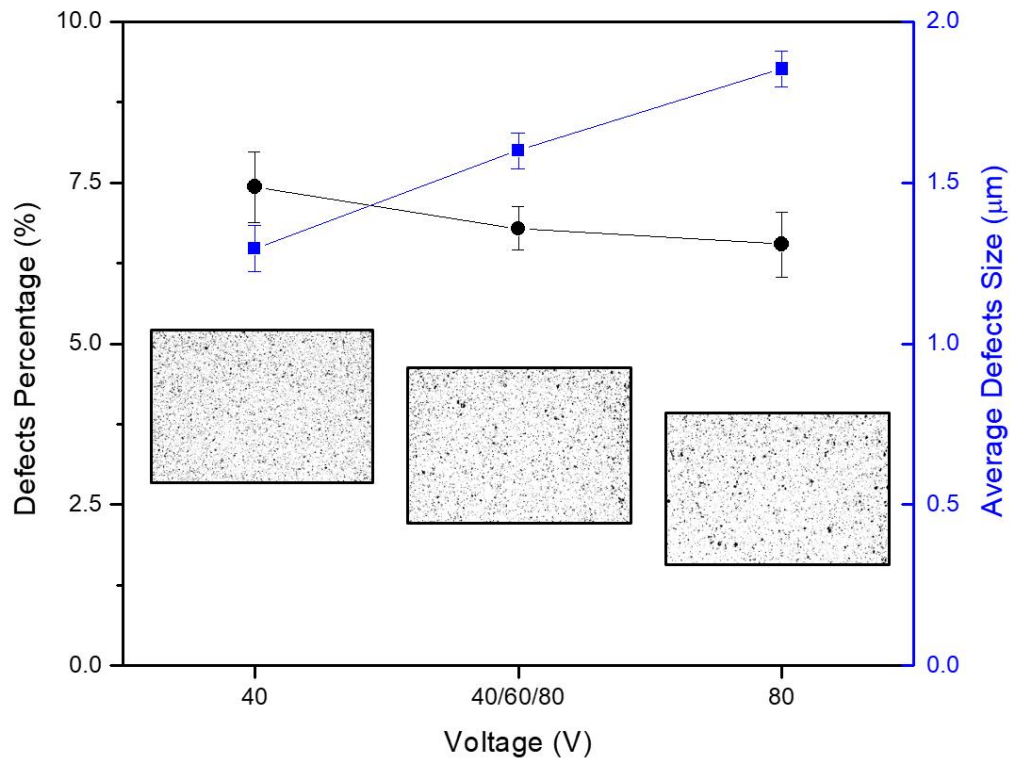

*Figure 2. Relationship between size and defects percentage with the bias voltage.*

A detailed analysis of the SEM micrographs show a honeycomb structure at CrN films surface, a similar behavior was found by some authors for CrN coatings deposited by arc-PVD (see Figure 3) [17,19]. The honeycomb structure is more defined in the 40/60/80Vb coating due to the gradual acceleration and momentum that suffer the ions and particles during the deposit process. On the other hand, flakes-like defects were observed that could arise from particles detachment of the chamber walls. Additionally, open void defects were found in all samples. This kind of defects are widely known and it is mainly due to a droplets removal or others particles deposited on the substrate surface. [20].

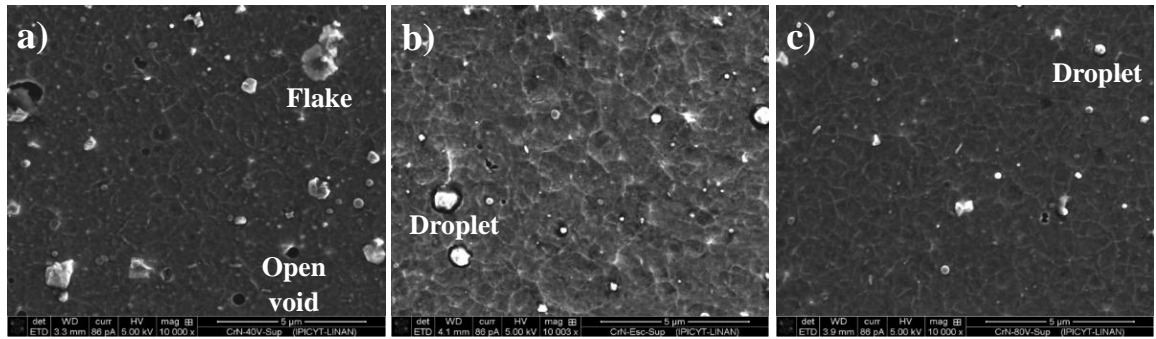

*Figure 3. SEM micrographs of CrN coatings: a) 40V$_b$, b) 40/60/80 V$_b$, and c) 80 V$_b$.*

*Mechanical properties*
Figure 3 show hardness (H), reduce elastic modulus (E), and the H$^3$/E$^2$ ratio of the CrN coatings. An increase 19.74 GPa to 25.38 GPa in hardness and 245.66 GPa to 293.41 GPa in the reduced elastic modulus was reached when bias voltage shift from 40 to 80 V. Values higher obtained to 80 V can be attributed to more dense microstructure and major lattice strain, which indicates compressive stresses that enhance the mechanical resistance of the system [21][22][23][24]. On the other hand, intermediate values of H and E were obtained with 40/60/80 V due to a combination of mechanical properties between 40Vb and 80V. Additionally, It is important to consider that the values found have been higher to reported by Tan [12], who obtained a hardness between 7 GPa to 15 Gpa with CrN coatings deposited with a graded bias voltage (from -20 V to -200 V with intervals of 5V, 10V, and 60V).

Besides, figure 3a shows the H$^3$/E$^2$ ratio of CrN coatings deposited at various voltages. This ratio represents the plastic deformation resistance of the coatings, which is related to the contact behavior of films. Higher values indicate the ability of the coating to absorb more energy. The behavior found is similar to the obtained with the hardness, in this case, the H$^3$/E$^2$ shows a growth from 113.45 GPa to 189.96 GPa with increasing of V, thus, this coating should be more rigid to the contact compare with the others.

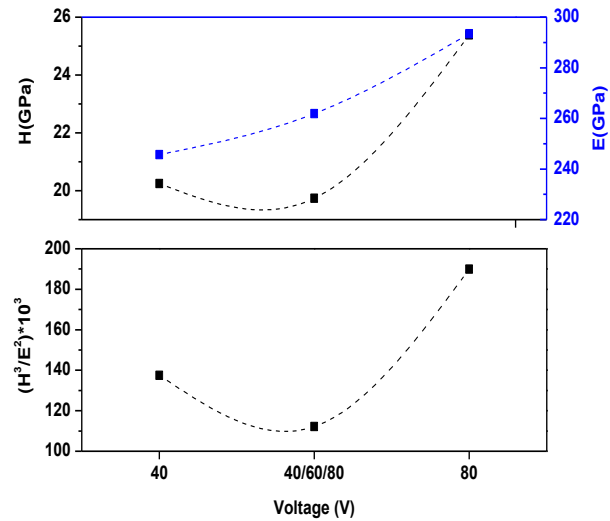

a)

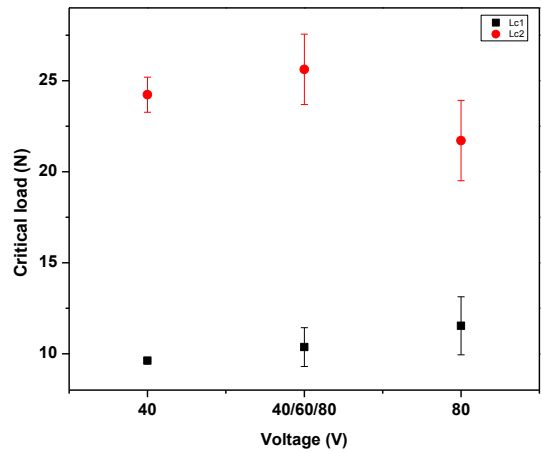

b)

*Figure 4. Effect of bias voltage on mechanical properties of CrN coatings: a) H, E and $H^3/E^2$; b) Critical loads obtained by scratch tester.*

Figure 4b shows the critical loads cohesive Lc1 (black) and adhesive Lc2 (red) related with resistances to initiation of cracks and to delamination. LC1 is influenced by $H^3/E^2$ ratio, thus, its behavior is similar to figure 3, where higher values were obtained with coatings deposited to 80 V. However, different behavior was found with LC2, in this case, the 40/60/80 V presents major interfacial resistance followed by coatings of 40 V and 80 V respectively. The phenomenon is related to the residual stresses observed by DRX, due to the high compressive stress in coatings with 80 V, the adhesion of the system of substrate/coating is reduces

[11][12]. On the other hand, the graded microstructure obtained with 40/60/80Vb diminishes the stress state produced by the mechanic contact.

Figure 5 shows the scratch tracks of the coatings. It was observed that at a load lower than LC1 the scratch track presents lateral cracks, corresponding to cohesive failures. When the load slightly overpasses LC1, recovery spallation at the border of the scratch crack and arc tensile cracks are formed, which are more marked by the increase of the load. After the Lc2 load, the scratch track shows buckling spallation. The buckling spallation is more severe 40/60/80V coatings likely by their honeycomb structure more pronounced.

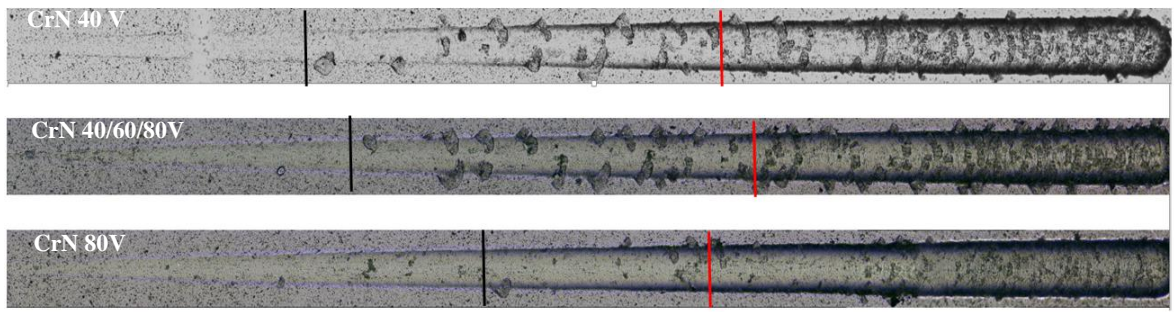

*Figure 5. Scratch tracks of the coatings. LC1-Black line; LC2- Red line*

Figure 6 shows the failure mechanism of the coatings. It can be seen that cohesive failures start with lateral cracks, which bordered droplets and they cross defects as voids and adhered weak particles. These cracks increase the depth when the load is rising up and they can interact with each other for originating recovery spallation at the border of the scratch crack (with values majors or equal to LC1), which act as cluster start point of buckling spallation. On the other hand, arc tensile cracks which are simultaneously generated with recovery spallation can aid to crack proliferation at end of the scratch track; this phenomenon is more notable when the honeycomb structure is more pronounced (40/60/80V). Additionally, the mechanism of failure depends on the $H^3/E^2$ ratio and the residual stress of films. For that reasons, at the scratch track end, the films deposited with 80 V has minor amount spallations.

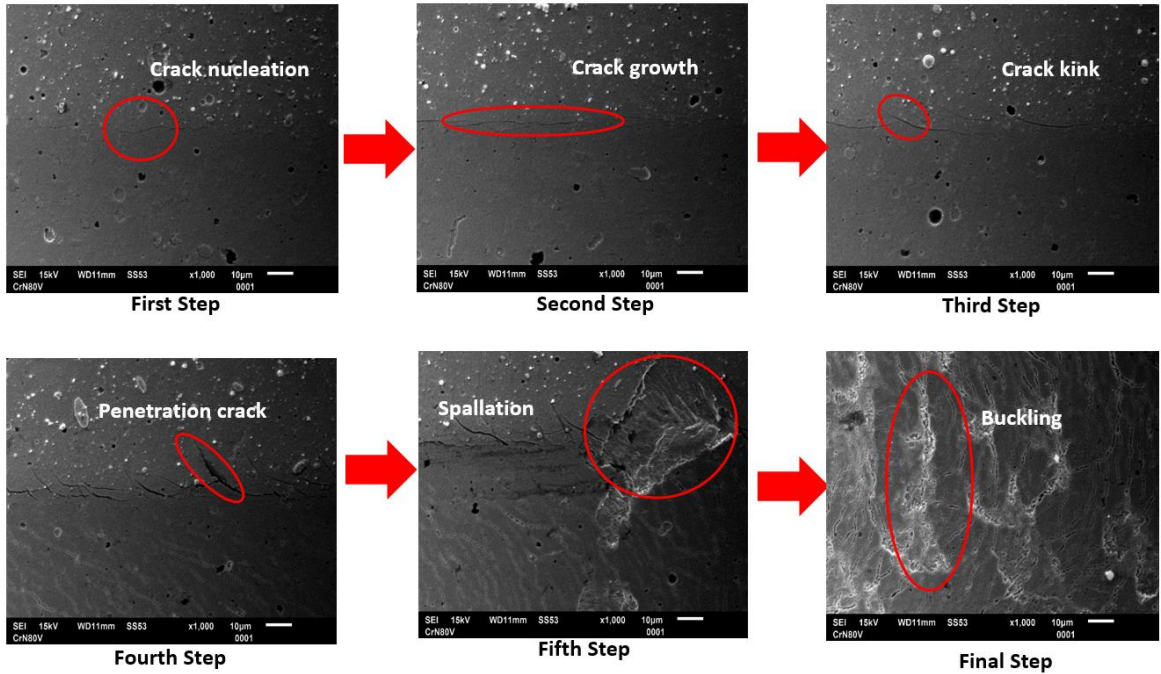

*Figure 6. Failure mechanism of the coatings (CrN 80V).*

*Electrochemical properties*

Figure 7 shows the potentiodynamic polarization curves of M2 steel without and with the coating, in presence of chlorides and at room temperature. The electrochemical parameters obtained by Tafel analysis are detailed in Table II.

*Table II. Electrochemical parameters*

| Material | Ecorr (mV) | Icorr (µA) | βa (mV/decade) | βc (mV/decade) | Protection efficiency (%) |
|---|---|---|---|---|---|
| Acero M2 | -619,87 ± 1,25 | 48,55 ± | 82,1 ± | 410,90 ± | N.A. |
| CrN 40V | 514,09 ± 7,47 | 0,05 ± 0,01 | 122.70 ± 0.01 | 188,10 ± 13.54 | 99.88 |
| CrN 40/60/80 V | 371,89 ± 13,14 | 0,25 ± 0,04 | 200,05 ± 37.54 | 185,15 ± 10.96 | 94.87 |
| CrN 80 V | -263,48 ± 4,70 | 2,48 ± 0,44 | 249,8 ± 62.49 | 115,35 ± 10.39 | 99.48 |

Where, *Ecorr*= corrosion potential, *Icorr*= corrosion current density, *βa*= anodic slope, *βc*= cathodic slope.

It was observed that M2 steel presents potential and corrosion current values of -619.87 mV and 48.55 µA respectively. In the anodic zone, its active region in the increases fasts with the applied potential until 0.3 V in which starting the formation of unstable and un-protective iron oxide layer, which allows the dissolution of the metal to continue taking place.

On the other hand, the CrN improves the corrosion resistance of M2 steel, and this is higher when reducing the bias voltage. That is showed with the increase in the corrosion potential and corrosion current lower values that were obtained with CrN 40Vb. In all cases coated steel exhibits a passive behavior, and the anodic current density increases slower than M2 steel, as shown by the larger value of the slopes of the anode region (Table II). Besides, all coatings have protection efficiency upper 90%.

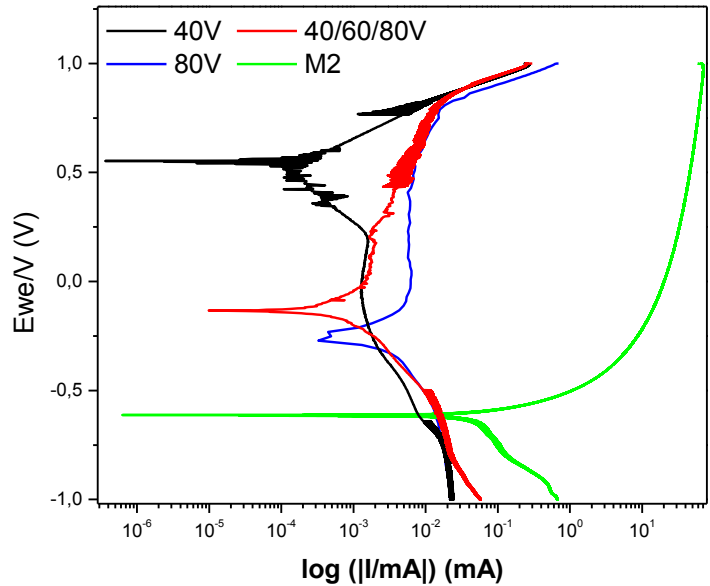

*Figure 7. Polarization curves of un-coating and coating steel.*

CrN coatings obtained to 40/60/80 V and 80 V have passive zones between potentials of -0.077 and 0.69 V derivated of the formation of protective oxide of $Cr_2O_3$. Besides, to potentials a little lower than transpassive potentials there are fluctuations in current due to an increase of dissolution points generated by pitting corrosion that the steel experiments when defects are removed of the coating. In films obtained to 40 V the pitting potential was not observed, only one active region could be identified. However, these films present the lower corrosion currents and the higher corrosion potentials, that indicated that this coating improves greater the corrosion resistance of M2 steel.

*Analysis of the surface corroded and corrosion mechanism*

Scanning electron microscope was used to analyze the surface of the specimens after the corrosion process. It was observed that the specimens are susceptible to medium chloride and present corrosion. A lower damage was observed in coated-steel. Figure 8 shows that M2 steel basically presents uniform corrosion, while that in coated-steel the localized corrosion is prevalent. In coated samples, the corrosion process is dependent on the voltage. As shown in figure 6 coatings deposited at 80 V had larger pitting and higher iron dissolution, while the corrosion process of CrN 40 V and CrN 40/60/80 V as was slower.

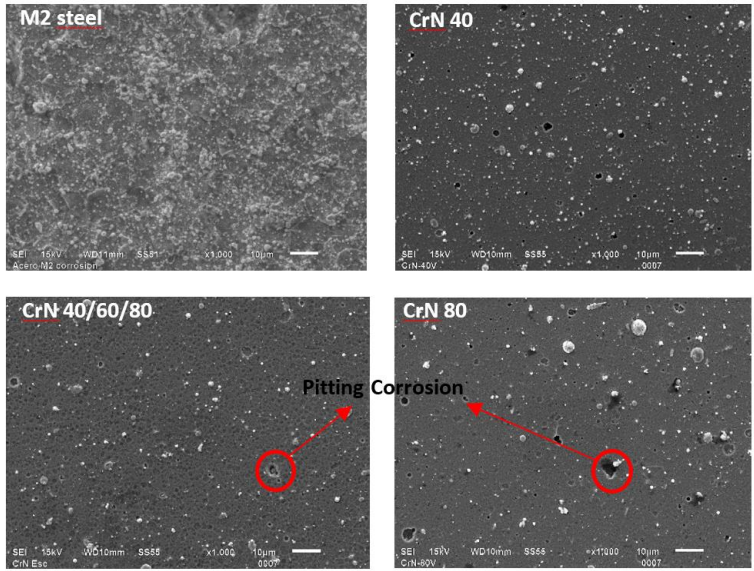

*Figure 8. SEM images of the surface after corrosion test.*

The corrosion phenomenon is triggered by the diffusion of the electrolyte in the porosities of the film, derived from defects formed during the deposition process. Pinholes can act as routes of diffusion of corrosive media towards the substrate. We found that the size of the droplets related to the corrosion mechanism, larger sizes generate a poor adhesion of the particle with the coating and can increase the residual stresses in the film favoring the separation of this inside the coating and subsequently the generation of spaces through which the electrolyte can penetrate, as shown in the figure 9. For that reason, CrN 80 V presented the lowest corrosion resistance.

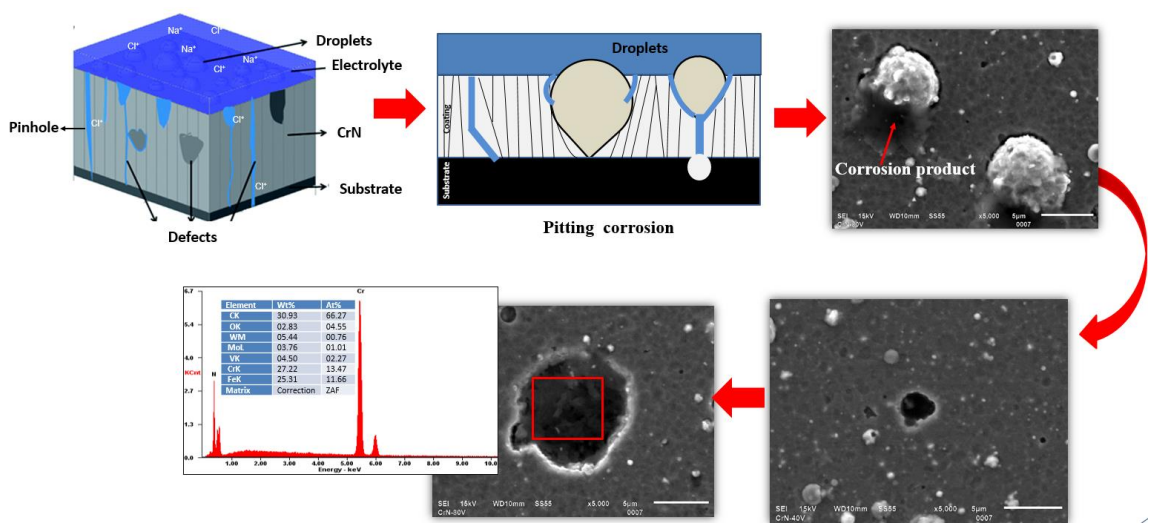

*Figure 9. Corrosion mechanism of the coatings.*

Besides, with the increase of the potential, the droplet detachment is produced and pore formation. The pores provide the direct diffusion ways for electrolyte and generate pitting corrosion.

## *Conclusions*

The effect of constant and graded bias voltage (40V, 80V, and 40/60/80V) in corrosion and mechanical properties were studied. We showed that the increase of bias voltage enhances the hardness and elastic modulus due to major lattice strain and compressive stresses in the film. However, the adhesion had an opposite behavior where CrN 40/60/80V and CrN 40V presented major critical load LC1.

The better adhesion observed with CrN 40/60/80V was attributed to the graded microstructure, which diminishes the stress state produced by the mechanic contact. In all cases, CrN coatings improved corrosion resistance of M2 steel with a protection efficiency upper 90%. The corrosion resistance was increased when the bias voltage was reduced to 40V, and the corrosion mechanism was major dominated by localized corrosion. The phenomenon was related to the defect sizes. A major size increase the corrosion current and shift down the potential, reducing the corrosion resistance.